\documentclass[%
aip,
amsmath,amssymb,
preprint,%
]{revtex4-1}

\usepackage{color}
\usepackage[english]{babel}
\usepackage{graphicx}% Include figure files
\usepackage{color}
\usepackage{physics}
\usepackage{dcolumn}% Align table columns on decimal point
\usepackage{bm}% bold math
\usepackage{subfigure}
\usepackage[utf8]{inputenc}
\usepackage[T1]{fontenc}
\usepackage{mathptmx}

\def\etal{{\textit{et al. }}}
\def\erfc{\mathrm{erfc}}
\def\erfi{\mathrm{erfi}}
\def\expe{\mathrm{e}^}
\def\ds{\displaystyle}
\begin{document}

\title{Neural network representability of fully ionized plasma fluid model closures}

\author{Romit~Maulik }
\affiliation{Argonne Leadership Computing Facility, Argonne National Laboratory, Lemont, IL 60439, USA}

\author{Nathan~A.~Garland}
\affiliation{Theoretical Division, Los Alamos National Laboratory, Los Alamos, NM 87545, USA}

\author{Xian-Zhu~Tang}
\affiliation{Theoretical Division, Los Alamos National Laboratory, Los Alamos, NM 87545, USA}

\author{Prasanna~Balaprakash }
\affiliation{Mathematics and Computer Science Division, Argonne National Laboratory, Lemont, IL 60439, USA}
\affiliation{Argonne Leadership Computing Facility, Argonne National Laboratory, Lemont, IL 60439, USA}

\date{\today}% It is always \today, today,

\begin{abstract}
The closure problem in fluid modeling is a well-known
challenge to modelers aiming to accurately describe their
system of interest. Over many years, analytic formulations
in a wide range of regimes have been presented but a practical, generalized fluid closure for magnetized plasmas remains an elusive goal. In this study, as a first step towards constructing a novel data based approach to this problem, we apply ever-maturing machine learning methods to assess the capability of neural network architectures to reproduce crucial physics inherent in popular magnetized plasma closures. We find encouraging results, indicating the applicability of neural networks to closure physics but also arrive at recommendations on how one should choose appropriate network architectures for given locality properties dictated by underlying physics of the plasma.

\end{abstract}

\maketitle

% \section{22 Nov Meeting notes/TO DO: }
% **Add baseline results compared to 'classical' machine learning methods - emphasise why the architecture chosen are useful, and fit for purpose.

% **What algorithm to use - and when? Combinations of certain limits? Come up with a hypothesis based on the data set. (Maybe?)

% **Mention analogy to efforts in fluids community, evidence that this is a reasonable approach.

% ** Motivation for fully-connected network? Not allowing global context over the domain, leads to poor result.

% ** Relate to Chenhao's paper, points of difference over that work. Where our H-P sections extends on that.

% ** Mention extrapolation and limits of that.

% ** CNN provides non-local support, but still has limits.

% ** One size network/surrogate does not fit all regimes that require closure. E.g. applying fully connected network to point-wise Braginskii problem, computational cost far too large.

% ** Mention by example the non-trivial nature of learning. E.g. the identity network of Romit's.

% ** Xianzhu to write some notes/thoughts on future ideas for this work

\section{Introduction}
In modeling and simulation of plasma physics dynamics many practitioners employ a fluid, or moment, model \cite{chen,braginskiireviewsofplasmaphysics1965,helander2005collisional}. The attraction of a fluid model lies in the generally simpler model description, and far reduced computational cost, when compared to high fidelity kinetic or particle based methods that describe a charged particle distribution function $f(\bm r,\bm v,t)$. To determine which of these two approaches to take, one must assess the trade-off between an averaged macroscopic description of a plasma that is computationally tractable, versus a fine kinetic microscopic description that may carry extreme computational cost. In the event of employing a fluid model, one must sufficiently truncate or close the infinite hierarchy of equations in the moments of $f(\bm r,\bm v,t)$.

While truncation at higher order may be done, it is more common to try to enforce some underlying physical properties of the problem to derive an approximate closure for higher order variables, such as heat flux, as a function of the lower order variables found in the model's conservation equations, such as density, velocity or energy. The physical properties used to derive closures can range from collisional to collisionless regimes, and can enforce local or non-local space-time relationships within the plasma \cite{held2004}. This being said, there are some scenarios where complex physics prevents a simple closure being assumed, and the question as to what closure to employ has a non-trivial answer. In this situation, some research communities, particularly fluid mechanics in recent years \cite{Yarlanki2012,CHANG2019559,SAN2018681}, have turned to machine learning to try to construct surrogate closure models that map the known macroscopic variables in a fluid model to the higher order moments that must be closed .

Machine and deep learning methods have long been used in fusion and plasma physics applications~\cite{Montes_2019,TangNature2019,Wroblewski_1997,Rastovic2012,Coccorese1994}, and in this study we seek to apply the methodology, inspired by the fluid mechanics community, of using machine learning to formulate and study surrogate models for the aforementioned magnetized plasma closure models that are crucial to fluid modeling. We note recent work of Ma \textit{et al.}~\cite{Ma_arxiv_2019} on the analysis of closure surrogates via neural networks for the Hammett-Perkins closure \cite{hammettphys.rev.lett.1990}. In this work we will also touch on the Hammett-Perkins non-local closure but also expand to include additional closures in different regimes relevant to fusion applications \cite{braginskiireviewsofplasmaphysics1965,guophys.rev.lett.2012}.

In seeking to learn surrogate closure models for use in a fluid model, we aim to provide some physical intuition of the problem \textit{a priori} before setting an arbitrary complex neural network on the problem. This style of approach has gained recent focus as physics-informed machine learning \cite{RAISSI2019686,WuPRF2018}, and we believe this problem can benefit greatly from this philosophy. With this in mind, the primary goal of this work was to verify the representability, and architecture style, of neural networks to sufficiently reproduce the often non-trivial physics that underpin the formulation of various analytic closure expressions.

We believe that benchmarking learned surrogates against well known analytic closures in important physical limits, and understanding the structure and limitations of the neural networks that formulate closure surrogates, is an instructive exercise. Particularly, it is useful to understand these limits before undertaking large-scale kinetic or particle simulations for data generation, assimilation, and subsequent closure learning in a haphazard manner. Further, from this study we can observe that, in the important physical limits grounding the closures studied, there is a benefit to tailoring a specific network architecture informed by the physics of the plasma regime each closure is designed for, rather than carelessly applying an unnecessarily complex general network architecture.

Our methodology and results are presented as follows. In Section~\ref{sec:one} we briefly review fluid models for magnetized plasmas and some of the closure methods used. Section~\ref{sec:two} outlines the learning method employed to formulate the closure surrogate neural networks, with Section~\ref{sec:three} detailing the network architectures employed for each surrogate. The results obtained from training the appropriate networks on each benchmark closure are discussed in Section~\ref{sec:four} followed by a summary of the key findings of this work in Section~\ref{sec:five}.

% \textbf{Key selling points of this paper:}\\
% 1. Local,non-local,point-wise,global - there's a benefit to tailoring the surrogate network informed by the physics\\
% 2. There are a number of physical limits of importance to plasma physics modeling. Here, we show in these limits ML surrogates are able to reproduce the crucial physics, especially in the non-local cases.  \\
% 3. Benchmarking analytic closures, with baked-in physics, is an important proof-of-concept step before performing large kinetic/particle simulations to assimilate data through a "black box" network to try to produce closures. \\
% 4. ... \\
% 5. ... \\

\section{Plasma fluid modeling}\label{sec:one}
The fundamental description of electron dynamics is carried by the electron distribution function, $f_e\left( \bm{r},\bm{v},t \right)$, in 7D phase space. While many techniques exist to resolve $f_e\left( \bm{r},\bm{v},t \right)$, the computational load can be enormous. The primary approach is solution of the kinetic equation for $f_e\left( \bm{r},\bm{v},t \right)$
\begin{align} \label{kinetic-eq}
\frac{\partial f}{\partial t} + \bm{v} \cdot \frac{\partial f}{\partial \bm{x}} + \frac{q}{m_e}\left( \bm{E} + \bm{v} \times \bm{B}  \right)\cdot \frac{\partial f}{\partial \bm{v}} = C(f),
\end{align}
where $C(f)$ is the collision operator.

As an alternative, a fluid description of the plasma is a popular approach that can simulate plasma properties in a more computationally efficient manner. Macroscopic fluid quantities, such density $n_e$, mean velocity $\bm{u}_e$, and pressure $\mathsf{P}_e$ are obtained via velocity moments of $f_e\left( \bm{r},\bm{v},t \right)$
\begin{gather}
n_e\left( \bm{r},t \right) = \int f_e\left( \bm{r},\bm{v},t \right) d^3\bm{v},\label{moment-n}\\
\bm{u}_e\left( \bm{r},t \right) = \frac{1}{n_e\left( \bm{r},t \right)}\int \bm{v}f_e\left( \bm{r},\bm{v},t \right) d^3\bm{v},\label{moment-u}\\
\mathsf{P}_e\left( \bm{r},t \right) = m_e\int \bm{w}\bm{w} f_e\left( \bm{r},\bm{v},t \right) d^3\bm{v},\label{moment-p}
\end{gather}
where $\bm{w}=\bm{v}-\bm{u}_e$ is the fluctuation velocity, $m_e$ is the electron mass, and the scalar pressure, $p_e$ often formed as a variable in continuity equations, can be extracted from the pressure tensor via $\ds \mathsf{P}_e = p_e\mathsf{I} + \mathsf{\Pi_e}$ where $\mathsf{I}$ is the identity tensor and $\mathsf{\Pi_e}$ is the viscosity tensor.

While fluid modeling is an attractive option to simulate plasmas, there are inherent challenges and disadvantages that this technique brings with it. Fundamentally, no exact knowledge of the velocity space distribution of electrons leads to an inherent inaccuracy in modeling discharges far from equilibrium, where a Maxwellian distribution function is often assumed. An extension of not knowing the velocity space distribution is the closure problem, where higher order moments, such as the viscosity tensor, $\mathsf{\Pi}_e$, or heat flux, $\bm{q}_e$, that appear in the derived continuity equations must be approximated in some way to close, or truncate, the system of equations to enable computation.

\subsection{Collisional limit: Braginskii closure}\label{bragclosure}
In his seminal review paper \cite{braginskiireviewsofplasmaphysics1965} Braginskii tackled the problem of showing convincingly that hydrodynamics reigns in a plasma made up of electrons and a single ion species moving through a strong magnetic field, where collisions occur at a rate that is comparable to the gyrofrequency. The well-known Braginskii~\cite{braginskiireviewsofplasmaphysics1965} equations for electron transport, dropping the subscript for $n$, $\bm{u}$, and $p$, can be written
\begin{gather}
\frac{dn}{dt} + n \nabla \cdot \bm{u} = 0, \label{eq:brag-n}\\
m_e n \frac{d\bm{u}}{dt} + \nabla p + \nabla \cdot \mathsf{\Pi} + e n \left(\bm{E} + \bm{u}\times \bm{B} \right) = \bm{F},\label{eq:brag-u}\\
\frac32 \frac{dp}{dt} + \frac52 p \nabla \cdot \bm{u} + \mathsf{\Pi} : \nabla \bm{u} + \nabla \cdot \bm{q} = W,\label{eq:brag-p}
\end{gather}

where $\ds \frac{d}{dt} = \frac{\partial}{\partial t} + \bm{u} \cdot \nabla$ is the commonly used convective or material derivative, $\mathsf{\Pi}$ is the viscosity tensor, $\bm{q}$ is the heat flux vector, $\bm{F}$ is a frictional force due to plasma resistivity and thermal forces, and $W$ is an energy transfer function due to collision with ions and work done by frictional forces.

Braginskii showed, through an asymptotic solution of the kinetic equation underlying equations (\ref{eq:brag-n}) - (\ref{eq:brag-p}), one can estimate the heat flux vector $\bm{q}$, the viscosity tensor $\mathsf{\Pi}$, the frictional force vector $\bm{F}$, and the energy exchange function $W$ in terms of the hydrodynamic observables.

In this study we seek to identify a network architecture that is capable of representing Braginskii's approximation to the closure given a range of macroscopic input profiles. To that end, a suitable litmus test is to assess the representability of one characteristic piece of the closure functional, namely the so-called gyroviscous stress tensor
\begin{align}
\mathsf{\Pi}^{\text{BG}}(n,\bm{u},T)(\bm{x}) = \frac{n(\bm{x})T(\bm{x})}{4 \norm{\bm{B}(\bm{x})}}\bigg(  \mathsf{\tilde\Pi} + \mathsf{\tilde\Pi}^T \bigg), \label{eq:BragStress}
\end{align}
where
\begin{gather}
\mathsf{\tilde\Pi} = \bm{\hat b}(\bm{x})\times (\nabla\bm{u}(\bm{x}) + \nabla\bm{u}^T(\bm{x}))\cdot (\mathsf{I} + 3\bm{\hat b}(\bm{x})\bm{\hat b}(\bm{x})),
\end{gather}
where $\bm{\hat{b}} = \bm{B}/\norm{\bm{B}}$ is the magnetic field unit vector and $\mathsf{I}$ is the 3 $\times$ 3 identity tensor.

The Braginskii formulation provides one benchmark limit for learned neural network architectures that could provide a surrogate closure model in the collisional magnetized plasma limit, with a local functional dependence. In contrast to this, the following sections now describe non-local closures in collisionless plasma closure limits.

\subsection{Collisionless limit: Hammett-Perkins closure}
Linear Landau damping is an inherently kinetic effect in weakly-collisional plasmas wherein fluid moments of the single-particle distribution function decay due to the development of increasingly-fine-scale structure in velocity space. These small-scale features are produced as a result of higher-velocity parts of the distribution function being transported by the streaming effect more rapidly than lower-velocity parts. A similar effect to Landau damping can be observed in sheared flows of conventional fluids \cite{Bedrossian_arxiv_2013}.

While the Landau damping effect is kinetic in nature, some aspects of the phenomenon can be mocked-up within a fluid model using an idea originally due to G. W. Hammett and F. W. Perkins.\,\cite{hammettphys.rev.lett.1990} Their basic idea was to develop a closure for the heat flux in terms of the temperature such that the decay rate predicted by linear Landau damping theory is reproduced exactly by the fluid model when linearized. In a single space dimension the Fourier-space expression for this closure is given by the simple formula
\begin{align}
    \hat{q}_k = -n_0\sqrt{\frac{8}{\pi}}v_{th} i\, \text{sgn}(k) \hat{T}_k.\label{hilbert_transform}
\end{align}
Here $n_0$ is the unperturbed plasma density and $v_{th}$ is the nominal plasma thermal speed. In a periodic domain the formula still makes sense provided the wave vector $k$ is suitably quantized.

Up to an unimportant proportionality factor the formula \eqref{hilbert_transform} says that the heat flux is related to the temperature by the famous Hilbert transform from signal analysis and elsewhere in plasma physics \cite{Morrison_2000,Heninger_2018}.\, In the signals context the transform is used, for example, to construct complex representations of signals that are analytic in the upper half-plane, while in the plasma context the transform is required to find the action-angle variables for the continuous spectrum of the linearized Vlasov-Poisson system. If $H$ denotes the Hilbert transform, and we neglect the unimportant constants, the closure relationship may be written $q(x) = H[T](x)$. There are a number of well-known formulas for the Hilbert transform including
\begin{align}
    H[T](x) = \frac{1}{\pi}\,P\int_{-\infty}^\infty\frac{T(s)}{x-s}\,ds,\label{PV_formula}
\end{align}
where $P$ denotes the principal value, and
\begin{align}
    H[T](x) = -\frac{1}{\pi}\lim_{\epsilon\rightarrow 0}\int_\epsilon^\infty \frac{T(x+s) - T(x-s)}{s}\,ds,
\end{align}
each of which explicitly displays the spatially-global nature of the closure. However where the Fourier space expression \eqref{hilbert_transform} needs only minor modification in a periodic domain (a quantization condition on $k$)  the previous two expressions require more significant changes. For instance the principal value formula \eqref{PV_formula} on a domain with period $2\pi$ becomes
\begin{align}
H[T](x) = \frac{1}{2\pi}\,P\int_0^{2\pi} T(s)\cot\left(\frac{x-s}{2}\right)\,ds.
\end{align}

In this study we seek to identify a network architecture that is capable of representing the Hammett-Perkins closure over a domain of temperature profiles sampled randomly from a Gaussian distribution with specified mean $\overline{T}(x) = \langle T(x)\rangle$ and covariance $\tilde{T}(x,x^\prime) = \langle(T(x)-\overline{T}(x))(T(x^\prime) - \overline{T}(x^\prime))\rangle$. That is, we treat the temperature as a Gaussian random field. A similar exercise was also carried out recently using a different training data generation method by Ma \textit{et al.}~\cite{Ma_arxiv_2019}. Like this earlier reference we will focus on learning the closure in configuration space in order to mask its trivial nature in Fourier space. In contrast to the earlier study our analysis will focus on the important issue of extrapolation errors inherent to artificial neural networks. We frame this extrapolation problem concretely as follows. If a neural network surrogate model for the Hammett-Perkins closure is trained using data sampled with mean $\overline{T}_1(x)$ and covariance $\tilde{T}_1(x,x^\prime)$ then how well does the learned closure perform on testing data sampled with mean $\overline{T}_2(x)$ and covariance $\tilde{T}_2(x,x^\prime)$? In any future application of neural networks to the problem of closure learning for plasma physics the limitations of the learning process set by extrapolation errors will be crucial to understand and manage.

\subsection{Collisionless limit: Guo-Tang closure}
In a collisionless, long $\lambda_\mathrm{mfp}$ magnetized plasma, $C(f)=0$, significant temperature anisotropy can develop. Further, if the plasma under consideration experiences an open magnetic field line, in contrast to the scenario for Hammett and Perkins closure \cite{hammettphys.rev.lett.1990}, it can be demonstrated that parallel heat flux can flow from low to high $T_\parallel$ in the vicinity of absorbing boundary walls, contradicting the classical thermal conduction description of heat flux i.e. $q \sim \kappa \nabla T$. In this scenario, a closure derived by Guo and Tang \cite{guophys.rev.lett.2012} has been demonstrated to sufficiently capture the behavior of the parallel heat flux in the presence of an absorbing boundary with open magnetic field line transport. In the Guo-Tang result closure expressions for both the electron and ion parallel heat flux components are presented, however in this study we solely examine the more physically interesting electron result.

\begin{figure}[!htb]
	\centering
	\includegraphics[width=0.6\linewidth]{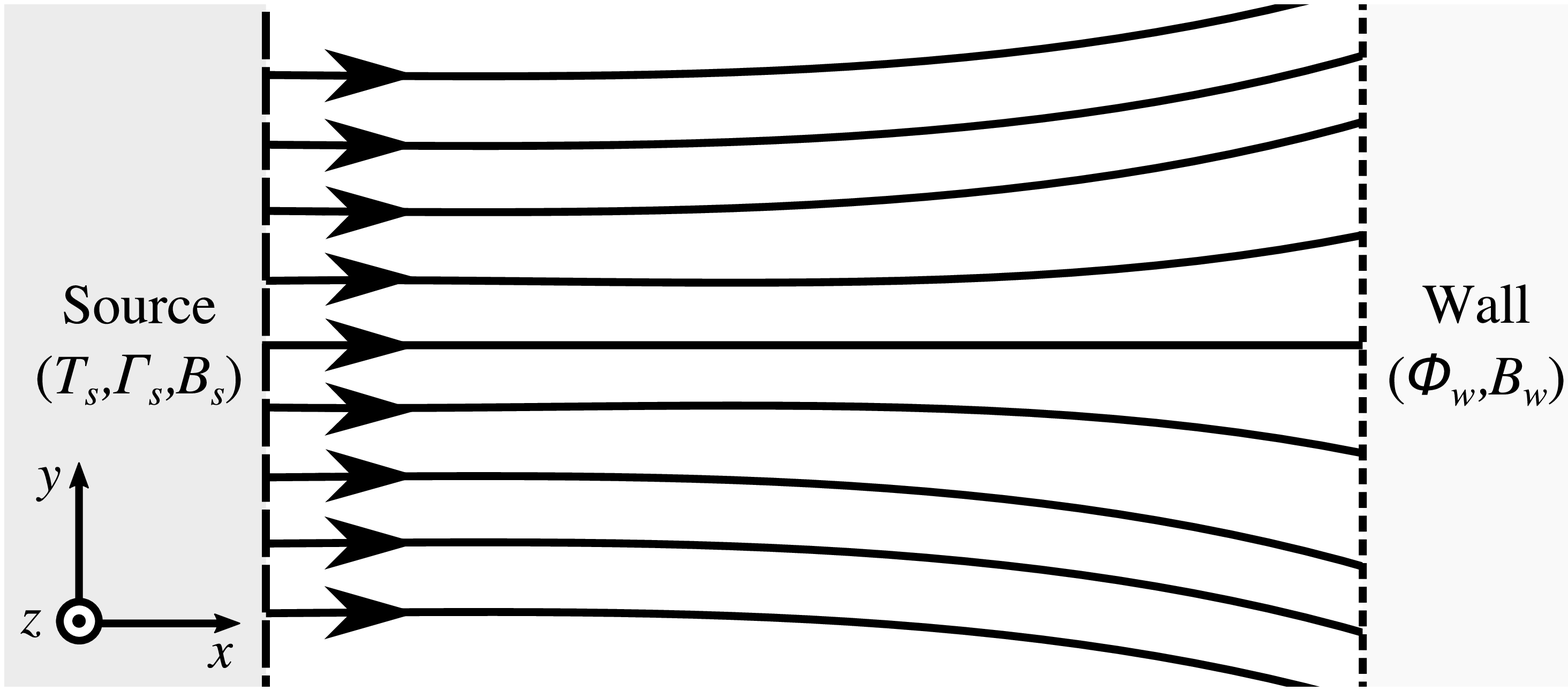}
	\caption{Geometry of Guo-Tang closure scenario \cite{guophys.rev.lett.2012}. A symmetric flux expander where the plasma is uniform in the z direction (out of plane) and symmetric about the source plane. Source parameters, $\Gamma_s$ and $T_s$, wall potential, $\phi_w$, and magnetic field magnitudes, $B_s$ and $B_w$, are employed in non-local closure. }
	\label{fig:GT_plasma}
\end{figure}

The Guo-Tang closure is applicable to electron transport within an open field line plasma bounded by absorbing walls, where a steady-state is achieved via balance with an upstream electron source plane with a Maxwellian distribution. Throughout the domain electrons experience an open magnetic field, $\bm{B}$, and ambipolar electric potential, $\phi$. A schematic example of this scenario is shown in Figure \ref{fig:GT_plasma}, indicating source flux, $\Gamma_s$, and temperature $T_s$, wall potential $\phi_w$, as well as magnetic field magnitude at the source, $B_s$, and wall $B_w$.

This closure relates the parallel heat flux, $\bm{q_\parallel}$, to contributions from parallel, $q_n$, and perpendicular, $q_s$, thermal energies
\begin{align} \label{eq:gt-parq}
\bm{q_\parallel} = \left( q_n + 2q_s \right )\bm{\hat{b}}.
\end{align}

The parallel, $q_n$, and perpendicular, $q_s$, contributions to the total heat flux are given by
\begin{gather}
q_n = \left \langle m_ev^3_{e\parallel}\right \rangle - 3n_eu_eT_{e\parallel}, \label{eq:qn}\\
q_s = \left \langle \mu_eBv_{e\parallel} \right \rangle - n_eu_eT_{e\perp}, \label{eq:qs}
\end{gather}
where the $n_eu_eT_e$ components represent convective fluxes available from the moment variables used in computation of the fluid model. The crux of the heat flux closure is in the evaluation of both $\left \langle m_ev^3_{e\parallel}\right \rangle$ and $\left \langle \mu_eBv_{e\parallel} \right \rangle$, given by
\begin{gather}
\frac12 \left \langle m_ev^3_{e\parallel}\right \rangle = \frac{\Gamma_s T_s}{2R^2}
 \left[  D + R\left( \frac32 - \frac{1}{R} + \frac{e\phi}{T_s} \right)\erfc(\sqrt{\tilde{\phi}_w}) \right . \notag\\
+ \left . C \left ( R\alpha \frac{e\phi}{T_s} + RG - G - \frac12  \right)\erfi(\sqrt{\alpha\tilde{\phi}_w})  \right],\label{eq:qn-close}\\
\left \langle \mu_eBv_{e\parallel} \right \rangle = \frac{\Gamma_s T_s}{2R^2}
\left [ -D + \erfc(\sqrt{\tilde{\phi}_w})
+ C \left ( G + \frac12  \right)\erfi(\sqrt{\alpha\tilde{\phi}_w}) \right]\label{eq:qs-close},
\end{gather}
where $R=B_s/B$, $e$ is the elementary charge, $\erfc()$ and $\erfi()$ are the complementary and imaginary error functions, and constants are defined as $\alpha = B_w/(B_s - B_w)$, $\tilde{\phi}_w= e\abs{\phi_w}/T_s$,
$C=\expe{-(1+\alpha)\tilde{\phi}_w}/\alpha^{3/2}$, $D=\expe{-\tilde{\phi}_w} \sqrt{ \tilde{\phi}_w/\pi}(1+\alpha)/\alpha $, $G=\alpha [ 3/2 + (1+\alpha)\tilde{\phi}_w]$.

In equations (\ref{eq:qn-close}) and (\ref{eq:qs-close}) the non-local dependence of the source and wall parameters on local quantities is explicit via $\Gamma_s$, $T_s$, $\alpha$, and most prominently $\phi_w$, which influences multiple terms.  The explicit non-local dependence in equations (\ref{eq:qn-close}) and (\ref{eq:qs-close}) echoes the observation of Hazeltine \cite{hazeltine1998}, in that the influence of the source and wall should enter explicitly in establishing a clear non-local description the parallel heat flux. It is exactly this non-local physics that warrants this closure model being chosen for this study, as it provides a useful benchmark limit, due to the presence of crucial non-local physics via source and boundary influence, for a learned neural network surrogate model.

\section{Details of the learning method}\label{sec:two}
In this section, we outline a workflow for machine learning the three types of closure introduced previously. The focus will be on bypassing the analytical expressions using neural network formulations. Individual networks for each closure will be designed according to the physical assumptions utilized in their analytical counterpart. In this way, we aim to provide a framework for generating closures in a physics-informed manner that simplifies the network designer's task for synthesizing closure surrogates. We note that the ultimate goal of this research is to generate closures that can deploy differing underlying assumptions according to localized deployment requirements.

\subsection{A surrogate Braginskii closure}
The underlying assumptions of the Braginskii closure rely on the specification of an analytical expression which is \emph{pointwise} in nature. To that end, our problem is formulated in a \emph{supervised} learning framework wherein the deep neural network is tasked with predicting the Braginskii closure at discrete locations on a grid locally. Training data is collected by utilizing the analytical form of the closure and framed in a feature-target representation where the features (or inputs) are given by the density $n(\bm{x})$, velocity $\bm{u}(\bm x)$, gradients of velocity $\nabla_{\bm x}\bm{u}(\bm x)$, and temperature $T(\bm{x})$ at a given point. The targets are given by the upper-diagonal components of $\mathsf{\Pi}^{\text{BG}}(n,\bm{u},T)(\bm{x})$. Note that the diagonal components are all zero and the target tensor is symmetric.

Training and testing data for the Braginskii closure surrogate was generated via
equation (\ref{eq:BragStress}) from randomly generated profiles for $(n,\bm u,T)$. These profiles were
specified to be continuous, smooth, and physical (i.e. $n, T>0$) via random mode sinusoids
\begin{gather} \label{eq:brag-ux-data}
n(x,y,z) = 0.5\sin(\alpha_1x)\sin(\alpha_2y)\sin(\alpha_3z) + 0.5,\\
u_x(x,y,z) = \sin(\alpha_4x),\\
u_y(x,y,z) = \cos(\alpha_5y),\\
u_z(x,y,z) = \sin(\alpha_6z),\\
T(x,y,z) = 0.5\cos(\alpha_7x)\sin(\alpha_8y)\cos(\alpha_9z) + 0.5,
\end{gather}
where random modes $0\leq \alpha_i \leq 3$ are generated from a uniform distribution.

% \subsubsection{Machine learning framework}
% For the purpose of learning, we utilize an artificial neural network that maps a relationship between the features and targets outlined above. Subsequently, this learning may be deployed at any region of a computational domain (provided the input features are available). For our map, we utilize a single-layer fully-connected neural network with 40 hidden-layer neurons. The advantage of this formulation is an ability to deploy on unstructured grids where local information is converted to pointwise estimates through the computation of gradients in a finite-difference or finite-volume sense.

% Hyperparameters related to the training of this framework are a batch-size of 128, i.e., the size of the subset of the entire data that is used for calculating the gradients for one updated of all the weights and biases (the trainable parameters) of the network and a learning-rate of 0.001 (which corresponds the the size of step taken in the direction of reducing error in prediction). The data generation for the training of this framework utilized 10 randomly generated velocity, density and temperature fields which had 50 grid points in each (of three) dimensions and their corresponding Braginskii closures. From each of these fields, 6000 samples were selected from random locations for a total of 60000 samples of inputs and outputs. A schematic of the closure is provided in Figure XX.

\subsection{A surrogate Hammett-Perkins closure}

The collisionless assumption of the Hammett-Perkins closure necessitates global context for pointwise predictions. This is reflected in the analytical expression defined in Fourier space with its global support. To incorporate this global context into a machine learning framework, we utilize a fully connected neural network that has global support. In other words a fully connected multi-layered perceptron is used to go from the inputs - given by the temperature profile to the targets - given by the heat-flux. The training data set for this assessment is generated from a temperature profile $\bar{T}$ that has a constant mean of 1.0 with a global context parameterized by a covariance given by
\begin{align}
    \tilde{T}(x,x') = \sigma^2 \exp(\frac{-2\sin(\pi(x-x')/T_p)^2}{L_c^2})
\end{align}
where $T_p$ is the periodicity of the samples and $L_c$ is the correlation length. Sampling from this multivariate distribution sequentially forms the inputs for our training data.

The mechanism for generating the training data also allowed for studying the effect of extrapolation by generating data from different covariance kernels. This will be demonstrated through visualization and quantitative assessments in the following sections.

% \subsubsection{Machine learning framework}

% The global context required for mimicking the Hammett-Perkins closure (on account of its definition in spectral space) motivates the use of a fully-connected network where the entire domain (i.e., in one dimension) is used as a direct input to the framework. The input layer is therefore the entire spatial field which is nonlinearly mapped through a hidden layer of size XX and rectified linear activation before being output through an output layer of the same dimension as the spatial grid.

\subsection{A surrogate Guo-Tang closure}

Training and testing data for the Guo-Tang closure is generated by evaluating non-local closure variables in equations (\ref{eq:qn-close}) and (\ref{eq:qs-close}) for a variety of randomly generated magnetic field and ambipolar potential profiles. Physical constraints require these profiles to be monotonic and decreasing \cite{guophys.rev.lett.2012}. To achieve profiles that are widely varying, to provide good training data, but also monotonic decreasing we use the exponential of a Gaussian cumulative distribution function, with random mean and variance. Over a domain $0\leq x \leq 100$ we generate input profiles
\begin{gather}
B(x) = B_s\exp(-\frac12 \left[ 1 + \mathrm{erf}\left(\frac{x-\hat X_1}{\sqrt 2 \sigma_1} \right)\right]), \label{eq:gt-B}\\
\phi(x) = \exp(-\frac12 \left[ 1 + \mathrm{erf}\left(\frac{x-\hat X_2}{\sqrt 2 \sigma_2} \right)\right]),\label{eq:gt-phi}
\end{gather}
where $B_s=2$, Gaussian means $ 10\leq \hat X_i \leq 90$ and variances $ 10 \leq \sigma^2_i \leq 60$ are randomly generated from a uniform distribution. Further, the temperature of source electrons $ 0.5\leq T_s \leq 2$ is randomly generated from a uniform distribution.

% \subsubsection{Machine learning framework}

% The collisionless Guo-Tang closure, similar to the Hammett-Perkins closure, requires some utilization of non-local context. Instead of parameterizing this relationship using a fully-connected network, we utilize a convolutional neural network (CNN) which uses localized filters to convolve inputs and gradually add non-local context through multiple layers. We note that CNNs generally require fewer training parameters and are widely used for data sets that may have anisotropic spatial correlations such as images.

\section{Machine learning architectures}\label{sec:three}

For the purpose of building surrogates for each closure hypothesis, we test three different types of machine learning frameworks given by fully connected, convolutional and locally connected artificial neural networks. A brief introduction is provided to all three frameworks and we demonstrate that each framework is associated with certain advantages (as well as disadvantages) and outline the need for surrogate closure development from data based on the underlying hypotheses in a physics-informed manner. In addition, we also comment on the choice of optimal hyperparameters for each of these frameworks and how these may be tied to the nature of non-locality in closure requirements.

\subsection{Fully connected neural network}

A fully connected neural network consists of a series of global operations that nonlinearly transform a set of inputs to a set of outputs given by input and output training data. In essence, all the inputs, which in our case are given by pointwise quantities, are flattened into a long vector of inputs and our outputs (also pointwise) are arranged in a similar fashion to have a point to point correlation between these vectors. Multiple fields correspond to multiple samples and therefore several fields of training data are needed to optimally train the free parameters of these networks. The nonlinear transformations in a fully connected network or otherwise are applied by multiple matrix operations on the flattened vector (which may or may not be used for transforming the size of the input vector) followed by subsequent nonlinear transformations of the transformed vector by \emph{activations} which are element-wise operations of nonlinear functions such as the sigmoid, hyperbolic tangent or rectified linear activations. An example of a sigmoidal activation is
\begin{align}
    \varphi(\beta)=\left(1+e^{-\beta}\right)^{-1},
\end{align}
where $\beta$ is a component of a vector. Similarly the hyperbolic tangent (also known as tanh) activation is given by
\begin{align}
    \varphi(\beta)=\left(e^{\beta}-e^{-\beta}\right) \cdot\left(e^{\beta}+e^{-\beta}\right)^{-1},
\end{align}
and the rectified linear activation (also denote ReLU) is given by
\begin{align}
    \varphi(\beta)=\max [0, \beta].
\end{align}
We note that ReLU is commonly used for its success in very deep networks as it prevents the saturation inherent in negative exponentiation in the sigmoid and tanh activations.

\subsection{Convolutional neural network}

Convolutional neural networks (CNNs) are a class of deep, feed-forward artificial  neural  networks that  are commonly  applied  to  analyzing  images. CNNs  assume  that the underlying images (and in our case field quantities) are stationary (i.e., statistics  of one part of the image are the same as other) and that a spatially local correlation exists between the image pixels. This is particularly important for modeling the artifacts introduced by lossy compression algorithms, whose effect on the underlying image is not dependent on the spatial location of the underlying image features. The convolutional layer is the core building block of a CNN and consists of filters whose size defines the extent of spatial locality assumed; each filter corresponds to a specific feature or pattern in the image. The convolution operation using these filters in each layer ensures stationarity and thus translational invariance. Several convolutional layers are stacked in such a way that the complex image features are learned hierarchically by composing together the features in previous layers. The CNN architectures are ideal for analyzing images, are more efficient to implement than fully connected models, and vastly reduce the number of parameters in the network, thus decreases the memory requirement and training time. However, it is generally observed that a greater amount of training data is needed for learning, particularly for spatial maps that need to be deployed pointwise. Greater details may be found in the seminal work of \cite{krizhevsky2012imagenet} which led to mainstream utilization of CNNs for image type problems.

\subsection{Locally connected neural network}

A locally connected neural network is a hybrid between CNN and FCNN architectures. It leverages non-local relationships in its input layer through feature engineering before using a fully connected network to map to the target values. More importantly, in the process of feature engineering, each grid point in the field is considered a sample. In this way, relationships are learnt pointwise without passing entire fields through the framework of large network architectures. Far simpler neural networks are obtained and training can be performed with fewer generation of fields since the number of samples is dramatically increased. The reader is directed to an example of such networks in \cite{maulik2017neural} for grid based learning problems. Note that for physics which are inherently global, this approach is limited as shall be demonstrated for the Hammett-Perkins closure.

\section{Results}\label{sec:four}

In this section we utilize statistical assessments for diagnosing the accuracy of machine learned predictions for the different types of architectures. These assessments are made through plots showing the probability distribution of machine learning responses compared to those of the truth. We also show scatter plots comparing predictions and truth. Each type of closure is assessed with the three types of neural network formulations we have outlined in the previous section both in terms of accuracy of predictions as well as the number of training parameters and the cost of training. We then make conclusions about how to select a particular architecture given \emph{a priori} understanding of the non-locality of a closure modeling scenario. All the networks assessed in the experiments of this section utilize a learning rate of 0.001 with a training duration determined by an early stopping criteria of 10. In other words, a section of the training data would be held out and not used for optimization (i.e., for validation) and training would be terminated if errors on this set were observed to be higher than the previous best error for more than 10 epochs. Batch sizes were varied for optimal throughput for all the experiments although results were observed to be relatively robust to different choices.

\subsection{Learning Braginskii}

We outline the results from using the machine learning model for predicting the Braginskii stress tensor in this subsection. Here, 6000 data points were utilized randomly from ten fields of $N_x \times N_y \times N_z = 50\times50\times50$ for the input and output quantities. We note that the predictions were for the off-diagonal components of a symmetric tensor (i.e., there were three only outputs from each query of the network). We shall provide three sets of results for the testing data (coming from a completely different field of inputs and outputs). Therefore, these plots describe the learning prowess of the networks on the data that it has not seen during trainable parameter optimization.

\begin{figure}[!htb]
\centering
\mbox{
\subfigure[]{\includegraphics[width=0.32\textwidth]{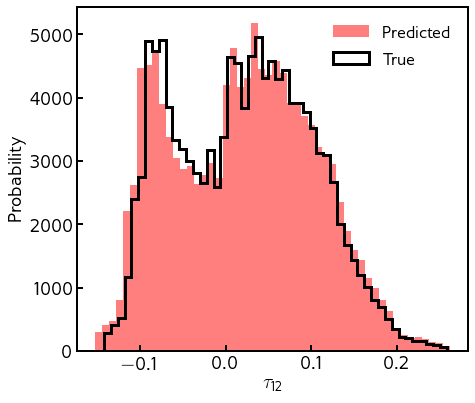}}
\subfigure[]{\includegraphics[width=0.32\textwidth]{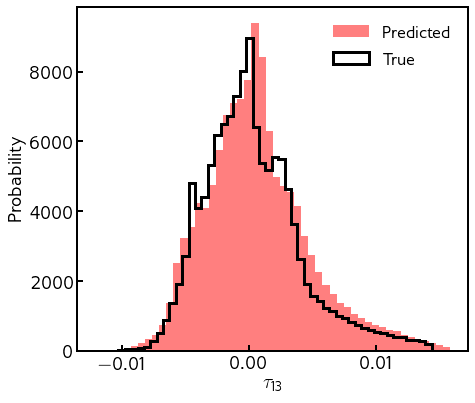}}
\subfigure[]{\includegraphics[width=0.32\textwidth]{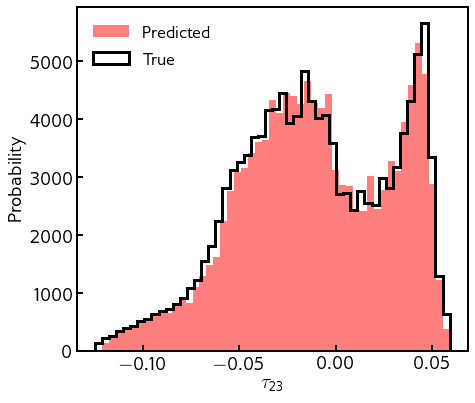}}
}
\caption{Probability density functions of upper diagonal Braginskii stress tensor elements (a) $\tau_{12}$, (b) $\tau_{13}$, (c) $\tau_{23}$ showing true and values predicted by the trained locally connected machine learning framework. Note that this data is a part of testing set showing the generalizability of the trained network.}
\label{fig:Brag_1}
\end{figure}

\begin{figure}[!htb]
\centering
\mbox{
\subfigure[]{\includegraphics[width=0.32\textwidth]{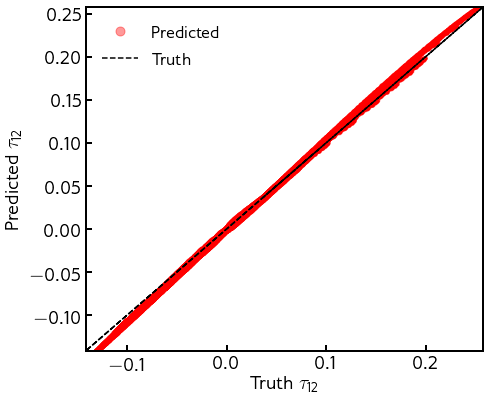}}
\subfigure[]{\includegraphics[width=0.32\textwidth]{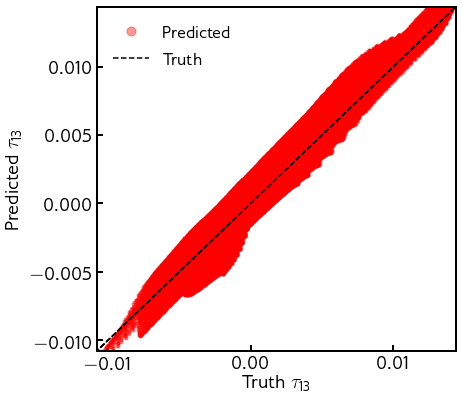}}
\subfigure[]{\includegraphics[width=0.32\textwidth]{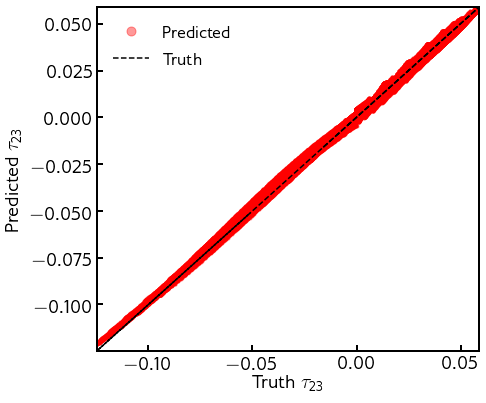}}
}
\caption{Scatter plots of upper diagonal Braginskii stress tensor elements (a)$\tau_{12}$, (b)$\tau_{13}$, (c)$\tau_{23}$ describing the accuracy of the locally connected machine learning framework. Deviations from the 45 degree line indicate erroneous predictions. Note that this data is a part of testing set showing the generalizability of the trained network.}
\label{fig:Brag_2}
\end{figure}

Figure \ref{fig:Brag_1} shows the probability density functions for the true and predicted values of the testing data set when using the locally connected neural network. This network architecture utilized each data point on the grid as a training sample and therefore built a map from 14 inputs (corresponding to the density, velocity, velocity gradient, and temperature information available from the grid) to 3 outputs corresponding to the closure terms. Response trends for each component are reproduced appropriately by this version of the surrogate formulation. Figure \ref{fig:Brag_2} shows the corresponding scatter plots for the same predictions. The 45 degree line represents the true values plotted against themselves. The predictions are seen to be quite close to the true values, represented by the predicted scatter points being close to the true line. The locally connected network utilized solely 1 hidden layer and with 40 neurons in each layer. We note that the coefficient of determination of this experiment was around $R^2=0.99$ indicating a very good fit.

\begin{figure}[!htb]
\centering
\mbox{
\subfigure[]{\includegraphics[width=0.32\textwidth]{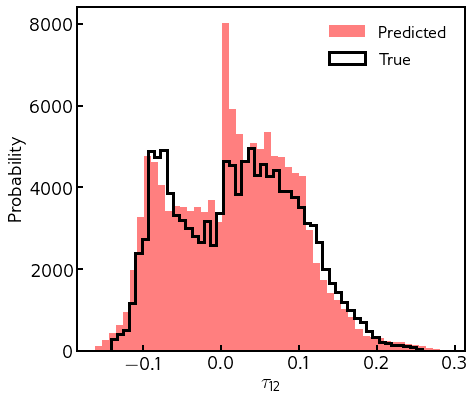}}
\subfigure[]{\includegraphics[width=0.32\textwidth]{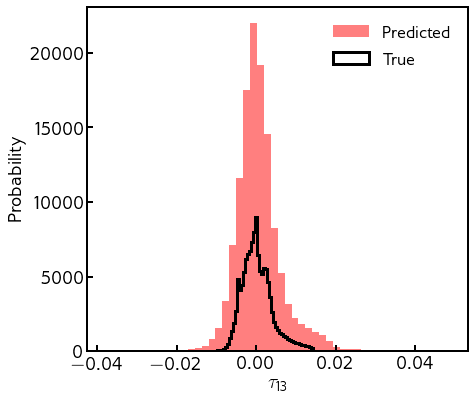}}
\subfigure[]{\includegraphics[width=0.32\textwidth]{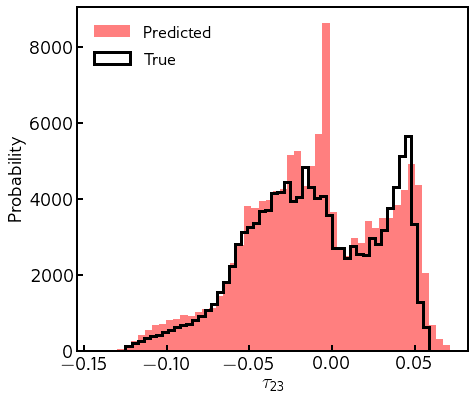}}
}
\caption{Probability density functions of upper diagonal Braginskii stress tensor elements (a) $\tau_{12}$, (b) $\tau_{13}$, (c) $\tau_{23}$ showing true and values predicted by the trained convolutional neural network learning framework. Note that this data is a part of testing set showing the generalizability of the trained network.}
\label{fig:Brag_3}
\end{figure}

\begin{figure}[!htb]
\centering
\mbox{
\subfigure[]{\includegraphics[width=0.32\textwidth]{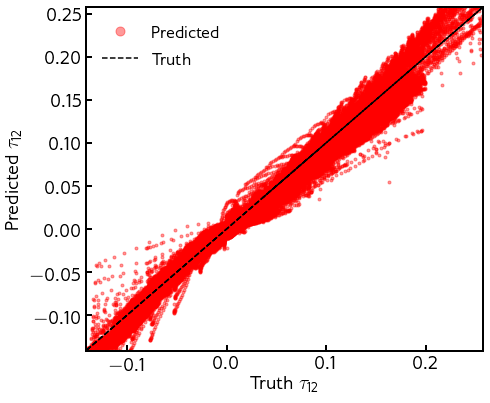}}
\subfigure[]{\includegraphics[width=0.32\textwidth]{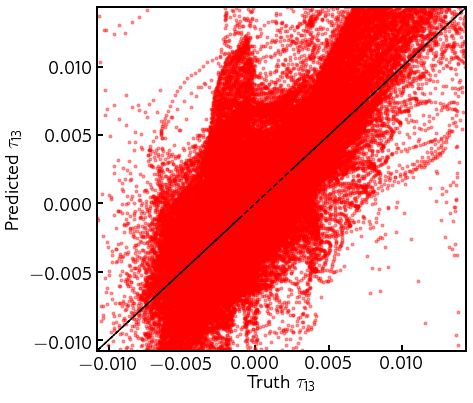}}
\subfigure[]{\includegraphics[width=0.32\textwidth]{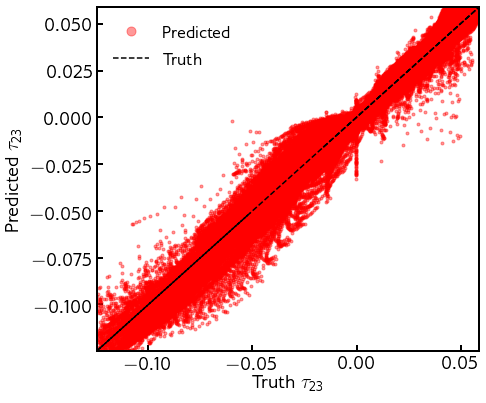}}
}
\caption{Scatter plots of upper diagonal Braginskii stress tensor elements (a) $\tau_{12}$, (b) $\tau_{13}$, (c) $\tau_{23}$ describing the accuracy of the convolutional neural network framework. Deviations from the 45 degree line indicate erroneous predictions. Note that this data is a part of testing set showing the generalizability of the trained network.}
\label{fig:Brag_4}
\end{figure}

Next we assess the ability of a CNN for the same prediction task. It is common knowledge that CNNs require more training data to learn effectively and we generate 100 fields of the input and target variables, representing a data set that is ten times larger than that used for the locally connected neural network. Training duration was also much reduced due to the greater number of floating point operations relative to one sample. We remind the reader that a data point for the CNN is an \emph{entire field}. The results for learning the Braginskii closure using a CNN are shown in terms of PDFs in Figure \ref{fig:Brag_3} and scatter plots in Figure \ref{fig:Brag_4} which show that learning may be improved (if compared to the locally connected network). We hypothesize that a greater amount of training data may improve the learning of the framework. The hyperparameters of the CNN for this assessment are given by 6 convolution layers with a filter sequence of [14,30,25,20,15,10,3]. The filter sequence controls the dimensionality of the number of channels of the data as it is being transformed through the network. To elaborate, the input data has 14 channels corresponding to all the input quantities available on the grid and the output tensor contains 3 channels that correspond to the closure terms. No pooling layers were used in this study and zero padding was used to preserve the dimension of the field during the forward pass through the framework. The $R^2$ of this particular experiment was slightly lower, but still very high, at 0.95.

\begin{figure}[!htb]
\centering
\mbox{
\subfigure[]{\includegraphics[width=0.32\textwidth]{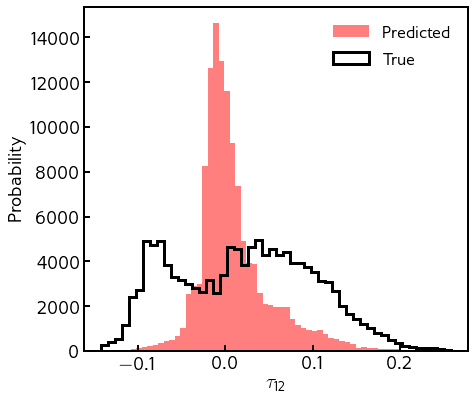}}
\subfigure[]{\includegraphics[width=0.32\textwidth]{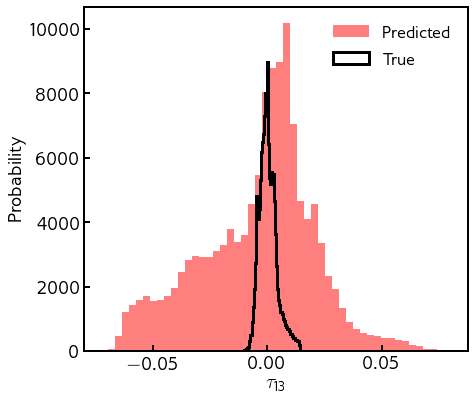}}
\subfigure[]{\includegraphics[width=0.32\textwidth]{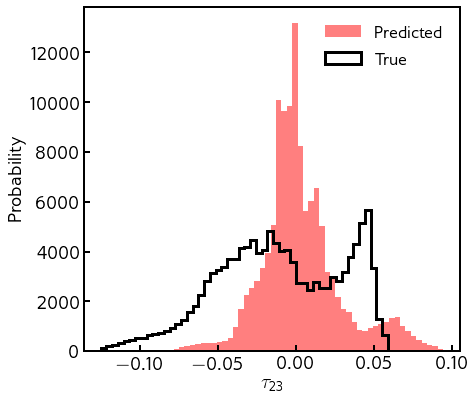}}
}
\caption{Probability density functions of upper diagonal Braginskii stress tensor elements (a) $\tau_{12}$, (b) $\tau_{13}$, (c) $\tau_{23}$ showing true and values predicted by the trained fully oonnected neural network learning framework. Note that this data is a part of testing set showing the generalizability of the trained network.}
\label{fig:Brag_5}
\end{figure}

\begin{figure}[!htb]
\centering
\mbox{
\subfigure[]{\includegraphics[width=0.32\textwidth]{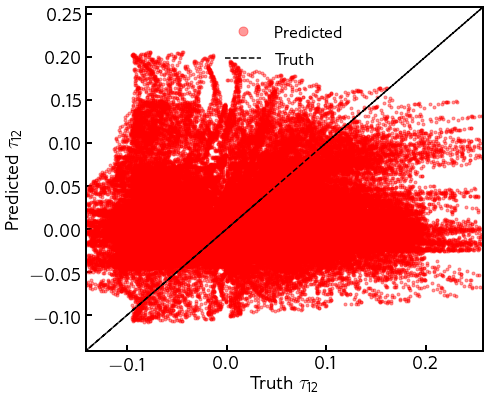}}
\subfigure[]{\includegraphics[width=0.32\textwidth]{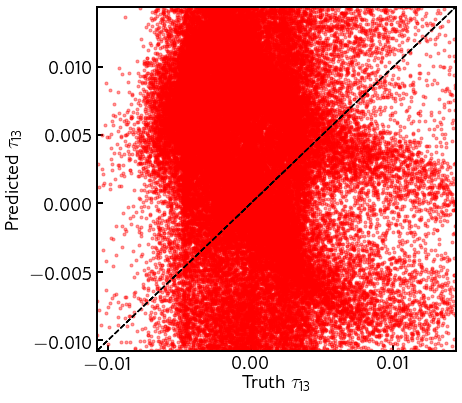}}
\subfigure[]{\includegraphics[width=0.32\textwidth]{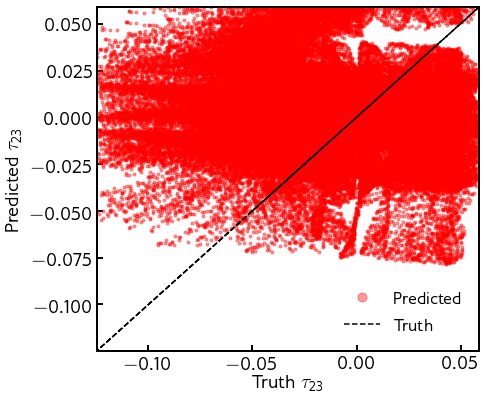}}
}
\caption{Scatter plots of upper diagonal Braginskii stress tensor elements (a) $\tau_{12}$, (b) $\tau_{13}$, (c) $\tau_{23}$ describing the accuracy of the fully connected neural network framework. Deviations from the 45 degree line indicate erroneous predictions. Note that this data is a part of testing set showing the generalizability of the trained network.}
\label{fig:Brag_6}
\end{figure}

Finally we show results from a fully connected network which interprets every field quantity at every points to a one dimensional vector and maps to the target tensor components also interpreted to be part of a large vector. Basically, our fully connected map is given by a nonlinear transformation from a space of dimension $N_x \times N_y \times N_z \times 14$ to $N_x \times N_y \times N_z \times 3$. One can note immediately, that the use of a fully connected network that connects every point in the field is computationally infeasible. Here, we demonstrate that training such a framework is also non-trivial as shown in Figures \ref{fig:Brag_5} (for the PDF) and Figures \ref{fig:Brag_6}. $R^2$ values of approximately 0.1 indicated the poor performance of this approach.

\subsection{Learning Guo-Tang}

\begin{figure}[!htb]
\centering
\mbox{
\subfigure[]{\includegraphics[width=0.42\textwidth]{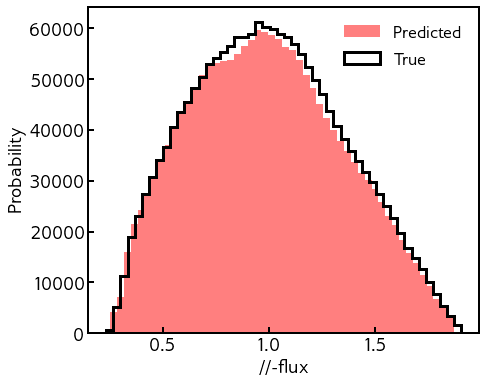}}
\subfigure[]{\includegraphics[width=0.42\textwidth]{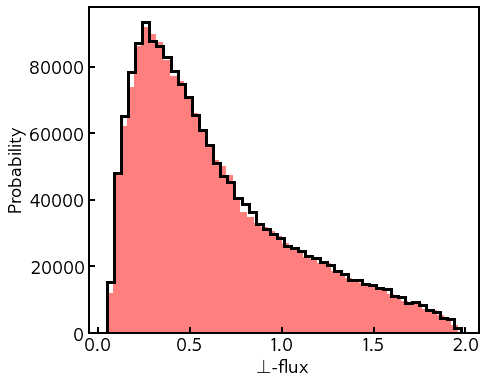}}
}
\\
\mbox{
\subfigure[]{\includegraphics[width=0.42\textwidth]{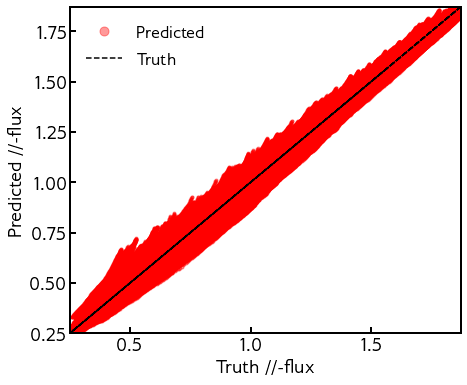}}
\subfigure[]{\includegraphics[width=0.42\textwidth]{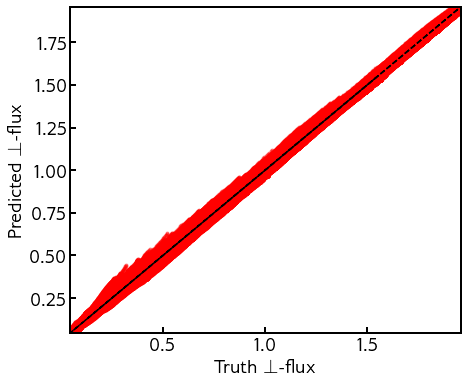}}
}
\\
\mbox{
\subfigure[]{\includegraphics[width=0.42\textwidth]{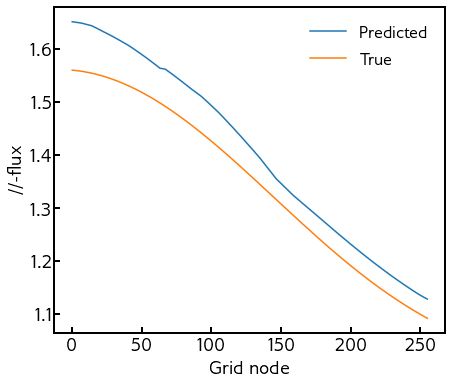}}
\subfigure[]{\includegraphics[width=0.42\textwidth]{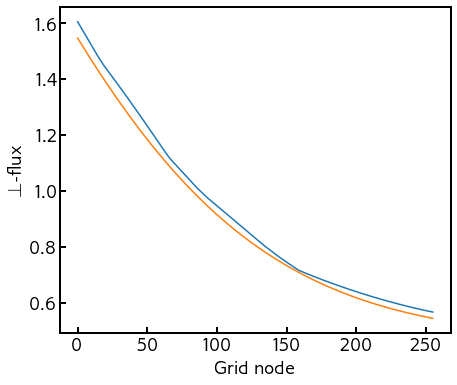}}
}
\caption{Machine learning results for learning the Guo-Tang closure for the locally connected machine learning framework. The top row shows PDFs of the true and predicted fluxes, the middle row shows the scatter plots for the same and the bottom row shows a sample field prediction by the trained frameworks. All assessments are on the test data set.}
\label{fig:GT_1}
\end{figure}

We proceed by assessing the ability of the machine learning frameworks for predicting the heat flux in the Guo-Tang framework. Figure \ref{fig:GT_1} shows the probability density functions of the true and predicted values of the parallel and perpendicular flux for the locally connected neural network. In addition they also show scatter plots and a sample realization of the true and predicted closure profile for one prediction. The locally connected neural network is able to obtain the right trends in the parallel and perpendicular fluxes from the point of view of PDFs and the scatter plots show a good agreement as well. Our locally connected neural network utilizes 4 hidden layers with 50 neurons in each layer to construct the local map between grid variables given by an input space of dimension 3, given by the magnetic field, ambipolar potential and source electron temperature, and output space of dimension 2, given by the two parallel and perpendicular components of the heat flux closure. We note that our grid is of dimension 256 but data at each location of the grid is considered independent of its neighbors for the purposes of training and deployment. $R^2$ values of 0.98 were observed for this experiment.

\begin{figure}[!htb]
\centering
\mbox{
\subfigure[]{\includegraphics[width=0.42\textwidth]{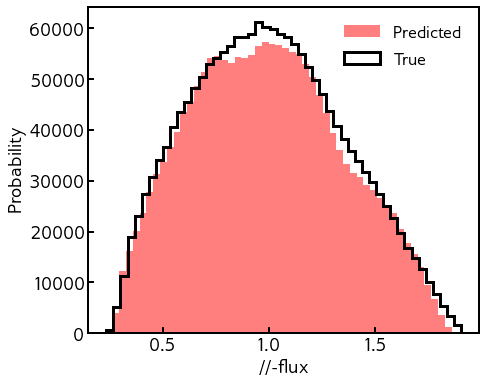}}
\subfigure[]{\includegraphics[width=0.42\textwidth]{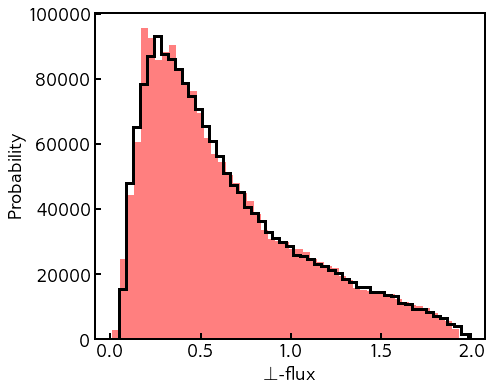}}
}
\\
\mbox{
\subfigure[]{\includegraphics[width=0.42\textwidth]{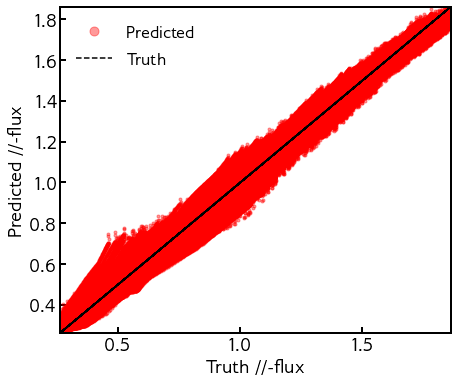}}
\subfigure[]{\includegraphics[width=0.42\textwidth]{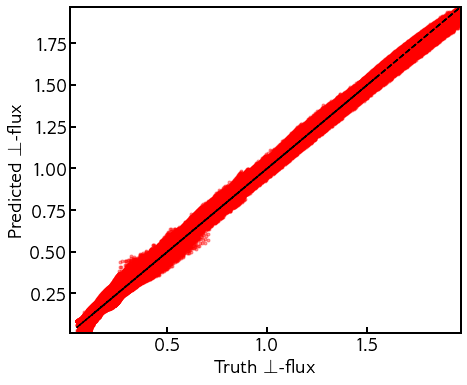}}
}
\\
\mbox{
\subfigure[]{\includegraphics[width=0.42\textwidth]{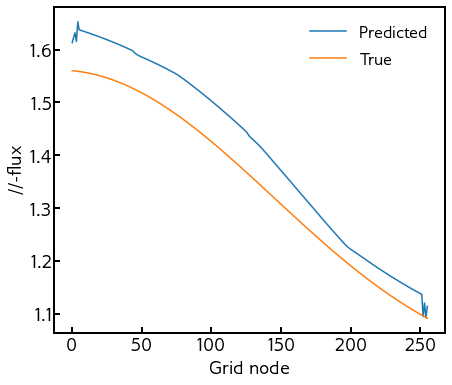}}
\subfigure[]{\includegraphics[width=0.42\textwidth]{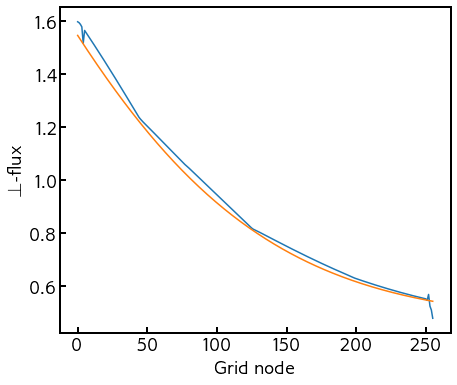}}
}
\caption{Machine learning results for learning the Guo-Tang closure for the convolutional neural network learning framework. The top row shows PDFs of the true and predicted fluxes, the middle row shows the scatter plots for the same and the bottom row shows a sample field prediction by the trained frameworks. All assessments are on the test data set.}
\label{fig:GT_2}
\end{figure}

Figure \ref{fig:GT_2} shows results from a deployment of the CNN framework for the same task. Similar trends are obtained for this assessment as well. However, we note that some boundary inaccuracies can be observed due to the strided nature of convolutional neural networks. These assessments utilized a zero-padding at the boundaries but more involved boundary conditions can also be embedded such as a periodic padding for each convolutional filter. Overall, the CNN is also able to learn the right nonlinear relationship for this closure. Our CNN architecture uses 6 convolutional layers with a filter sequence of [2,30,25,20,15,10,3] and converged to an $R^2$ value of 0.98.

\begin{figure}[!htb]
\centering
\mbox{
\subfigure[]{\includegraphics[width=0.42\textwidth]{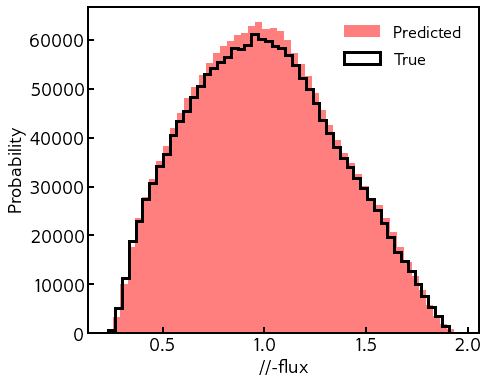}}
\subfigure[]{\includegraphics[width=0.42\textwidth]{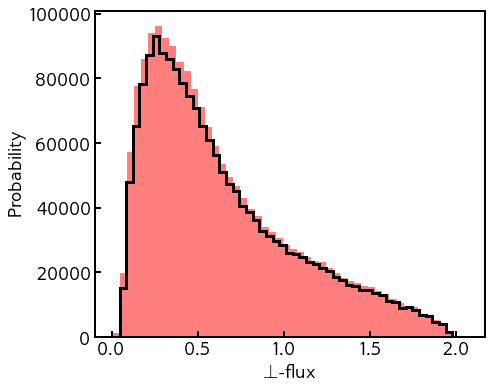}}
}
\\
\mbox{
\subfigure[]{\includegraphics[width=0.42\textwidth]{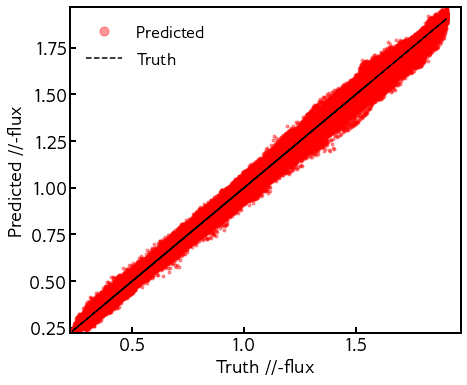}}
\subfigure[]{\includegraphics[width=0.42\textwidth]{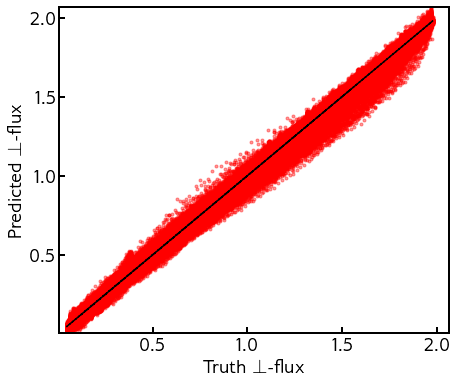}}
}
\\
\mbox{
\subfigure[]{\includegraphics[width=0.42\textwidth]{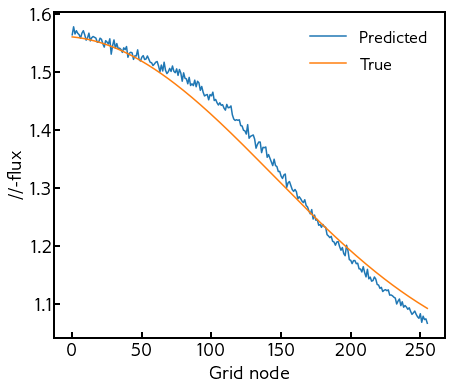}}
\subfigure[]{\includegraphics[width=0.42\textwidth]{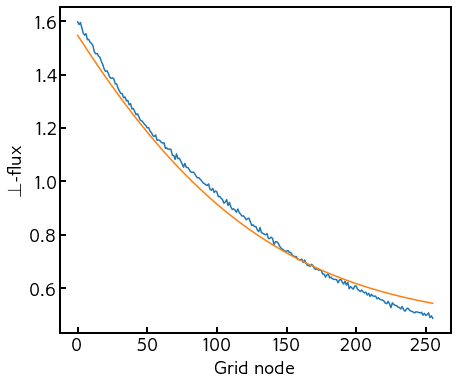}}
}
\caption{Machine learning results for learning the Guo-Tang closure for the fully connected machine learning framework. The top row shows PDFs of the true and predicted fluxes, the middle row shows the scatter plots for the same and the bottom row shows a sample field prediction by the trained frameworks. All assessments are on the test data set.}
\label{fig:GT_3}
\end{figure}

Figure \ref{fig:GT_3} shows results from a deployment of the fully-connected neural network. Good results are obtained for this assessment although field values can be observed to be slightly noisy. This noise is a result of the fully connected network being sensitive to perturbations across the entire domain, hence small fluctuations of input variables through the domain can lead to jitter in the observed output quantities through the entire domain as well. However, despite the presence of noise, the predictions from the fully connected framework show less deviation (particularly in the case of the parallel flux) from the true values. This might suggest a suitable use of the framework along with a low-pass spatial kernel used for postprocessing the output. We note that our neural network utilizes 4 hidden layers with 50 neurons in each layer to construct the local map between grid variables given by an input space of dimension 256x3 and output space of dimension 256x2 with converged $R^2$ values of 0.98.

\subsection{Learning Hammett-Perkins}

\begin{figure}[!htb]
\centering
\mbox{
\subfigure[]{\includegraphics[width=0.32\textwidth]{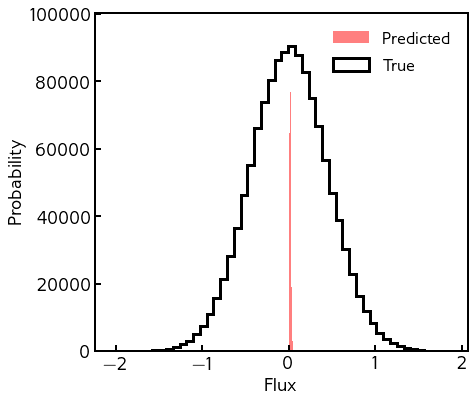}}
\subfigure[]{\includegraphics[width=0.32\textwidth]{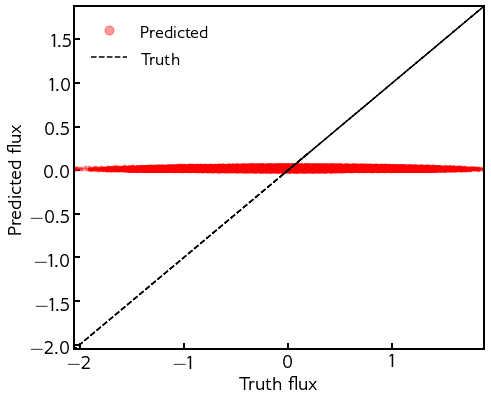}}
\subfigure[]{\includegraphics[width=0.32\textwidth]{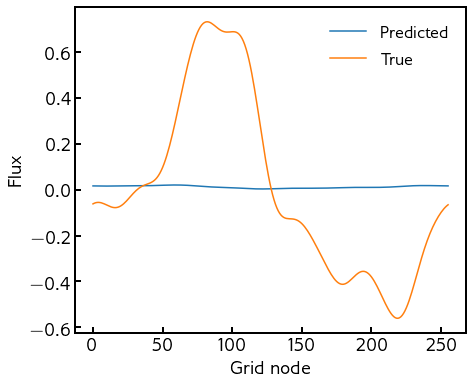}}
}
\caption{Results from machine learning the Hammett-Perkins closure with probability density functions (left), scatter plots (middle) and a sample prediction visualized against the truth (right) for the locally connected neural network. These are generated for the testing data set.}
\label{fig:HP_1}
\end{figure}

We perform a third analysis (similar to the previous two) for the Hammett-Perkins closure with assessments for the locally connected neural network shown in Figure \ref{fig:HP_1}. The global nature of Hammett-Perkins is immediately discernible through the performance of the trained local framework which fails to capture the right trends of the output almost entirely. This was observed for multiple different choices of the hyperparameters of the network. The results here are shown for 2 hidden layers and 30 neurons in each layer and attempt to map from a 1 dimensional input of temperature profile to a 1 dimensional output of the heat flux.

\begin{figure}[!htb]
\centering
\mbox{
\subfigure[]{\includegraphics[width=0.32\textwidth]{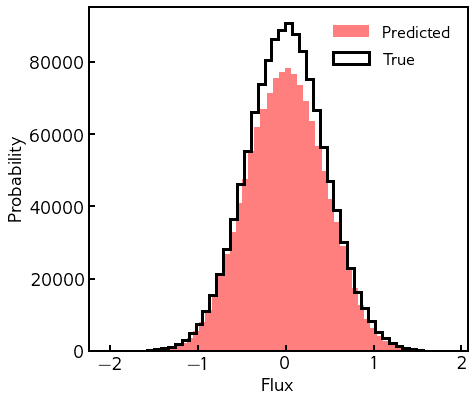}}
\subfigure[]{\includegraphics[width=0.32\textwidth]{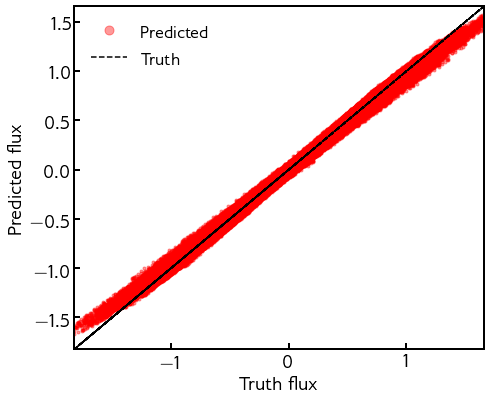}}
\subfigure[]{\includegraphics[width=0.32\textwidth]{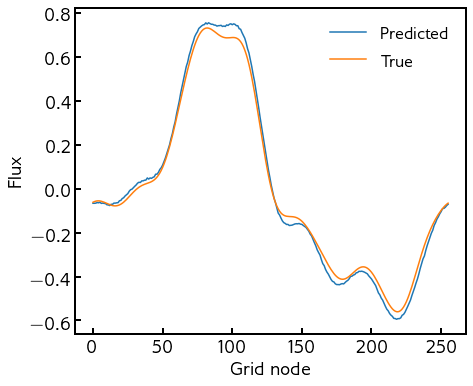}}
}
\caption{Results from machine learning the Hammett-Perkins closure with probability density functions (left), scatter plots (middle) and a sample prediction visualized against the truth (right) for the fully connected neural network. These are generated for the testing data set.}
\label{fig:HP_3}
\end{figure}

In contrast, Figure \ref{fig:HP_3} shows the results of a CNN with a finite non-locality where it is observed that some trends are learnt although the accuracy is marginal. The size of the kernel and the number of CNN layers are instrumental in the accuracy of the predictions. Essentially, an increase in the number of layers with a local stencil increases global influence and may be assumed to improve accuracy. However, we note that the perfectly global nature of Hammett-Perkins implies that nothing short of a fully connected network can obtain optimal learning. These results are shown for a filter sequence of [1,30,25,20,15,10,1] with only one input and one output and obtained $R^2$ values of 0.94.

\begin{figure}[!htb]
\centering
\mbox{
\subfigure[]{\includegraphics[width=0.32\textwidth]{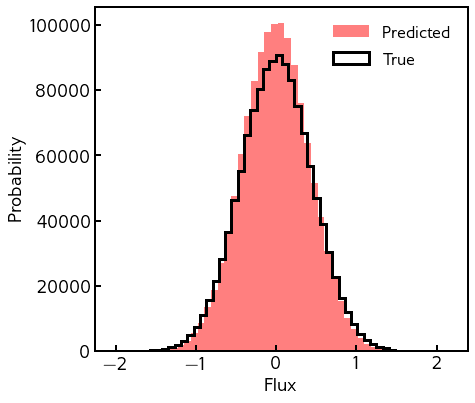}}
\subfigure[]{\includegraphics[width=0.32\textwidth]{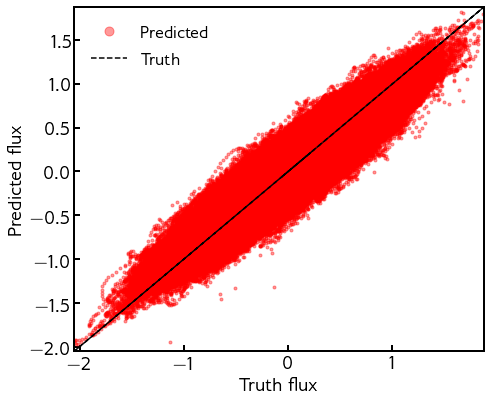}}
\subfigure[]{\includegraphics[width=0.32\textwidth]{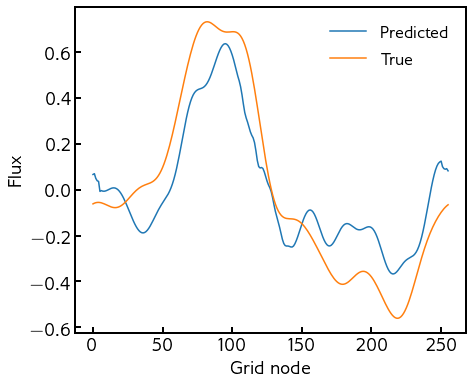}}
}
\caption{Results from machine learning the Hammett-Perkins closure with probability density functions (left), scatter plots (middle) and a sample prediction visualized against the truth (right) for the convolutional neural network. These are generated for the testing data set.}
\label{fig:HP_2}
\end{figure}

The results of the fully connected network are shown in Figure \ref{fig:HP_3} where a much improved performance is obtained by the machine learning architecture. The fully connected neural network mimics the global nature of the Fourier-space transformation and is thus able to recover the trends most accurately. A conclusion from this set of experiments is that for prediction tasks where the interaction between inputs is perfectly global, a fully connected network is optimally suited. However, due to the greater computational expense in training and deployment of these networks, one may also utilize a sufficiently deep CNN. The fully connected networks were able to obtain large $R^2$ values of 0.99 and are optimally suited for this type of closure requirement. The architecture for this task was given by 2 hidden layers with 30 neurons mapping inputs from 256 dimensional inputs to 256 dimensional outputs.

To summarize the differing complexities of the proposed network architectures, we outline the number of trainable parameters for each closure learning in Table \ref{Table_Parameters}.

\begin{table}[]
\centering
\small
\begin{tabular}{c|c|c|c|}
\cline{2-4}
\multicolumn{1}{l|}{}                    & \multicolumn{3}{c|}{\textbf{Network architectures}}                                       \\ \hline
\multicolumn{1}{|c|}{\textbf{Closure type}}       & Fully connected & Convolutional & Locally connected \\ \hline
\multicolumn{1}{|c|}{Braginskii} & 64,128,750 & 58,133 & 723 \\ \hline
\multicolumn{1}{|c|}{Guo-Tang} & 72,212     & 5,532  & 7,952 \\ \hline
\multicolumn{1}{|c|}{Hammett-Perkins} & 16,576     & 5,321  & 1,021 \\ \hline
\end{tabular}
\caption{The number of trainable parameters for each neural network architecture for the three different closure scenarios. It is apparent that the fully connected neural networks require a large number of parameters in comparison to the CNN or locally connected networks.}
\label{Table_Parameters}
\end{table}

\begin{table}[]
\centering
\small
\begin{tabular}{c|c|c|c|}
\cline{2-4}
\multicolumn{1}{l|}{}                    & \multicolumn{3}{c|}{\textbf{Network architectures}}                                       \\ \hline
\multicolumn{1}{|c|}{\textbf{Closure type}}       & Fully connected & Convolutional & Locally connected \\ \hline
\multicolumn{1}{|c|}{Braginskii} & $1.10\times 10^{-2}$  & $1.66\times 10^{-4}$ &  $2.40\times 10^{-5}$\\ \hline
\multicolumn{1}{|c|}{Guo-Tang} & $2.93\times 10^{-4}$      & $1.09\times 10^{-3}$  & $4.84\times 10^{-4}$ \\ \hline
\multicolumn{1}{|c|}{Hammett-Perkins} &  $7.54\times 10^{-4}$    &  $2.71\times 10^{-2}$ &  $1.97\times 10^{-1}$ \\ \hline
\end{tabular}
\caption{The mean-squared errors for each neural network architecture for the three different closure scenarios.}
\label{MSES}
\end{table}

\subsection{A note on extrapolation}

In the previous sections, we have demonstrated the applicability of learned surrogates when subject to the domain of the training data. In practice, one may have a lot of data to learn from but that data may not have a broad span of a system's parameter space to generate training data; this may be considered as a balance between \textit{big data}, i.e. a large quantity, and \textit{broad data}, a large \textit{and} statistically meaningful quantity of data. Thus, predictions may have to be extrapolated based on what was learnt from the available training domain. Before presenting outcome of the closure surrogates in extrapolating regimes, we wish to note that extrapolating outside the training domain using a learned neural network architecture is a non-trivial task that requires careful application. A simple demonstrative example of the unreliability of trying to emulate even the most simple of functions, the identity $f(x)=x$, was presented recently \cite{gin2019deep}. While an exact analytic solution of network coefficients can be found for a simple network architecture like the identify function, to allow accurate extrapolation outside a training domain, Gin \etal highlight the sensitivity of learned network coefficients and thus the reliability of predictions outside the training domain even for such a simple problem. We believe this is an important lesson to keep in mind when applying neural network methods to real problems, and emphasizes that a network may learn to mimic behaviour of a set of training data but cannot reliably learn the underlying general functional relationships relating input and output quantities.

\subsection{Testing for extrapolation}

\begin{figure}[!htb]
\centering
\mbox{
\subfigure[]{\includegraphics[width=0.32\textwidth]{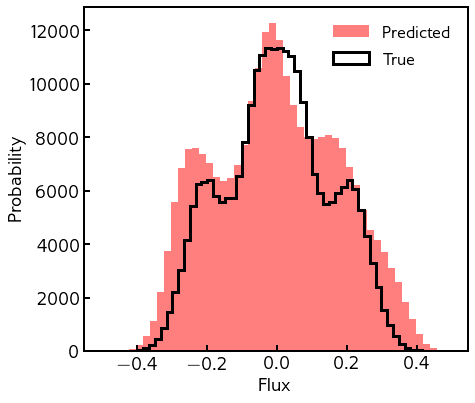}}
\subfigure[]{\includegraphics[width=0.32\textwidth]{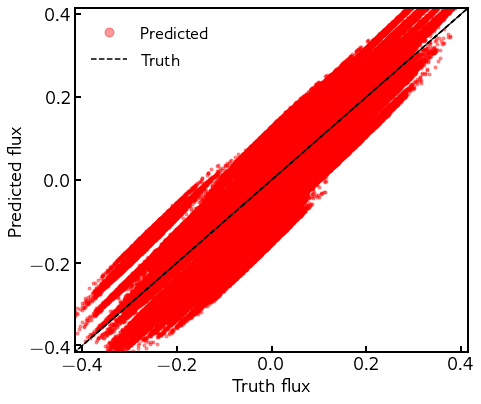}}
\subfigure[]{\includegraphics[width=0.32\textwidth]{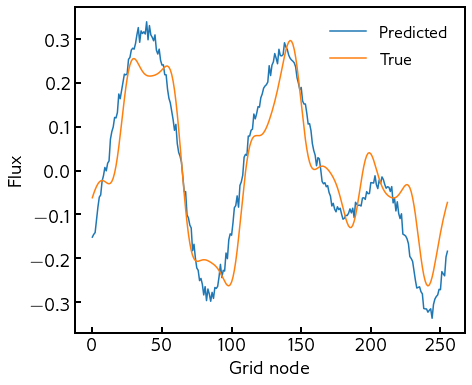}}
}
\caption{Results from machine learning the Hammett-Perkins closure with probability density functions (left), scatter plots (middle) and a sample prediction visualized against the truth (right) for a fully connected neural network. These are generated for the testing data set from a different underlying distribution. Results of extrapolation outside the space of trained input data can be clearly observed.}
\label{fig:HP_4}
\end{figure}

One of the key challenges of machine learning is to detect extrapolation. Machine learning frameworks are notoriously unreliable in extrapolating regimes as mentioned previously. Detecting extrapolation is useful for determining if more data should be generated and the framework should be retrained using this new data before deployment. For example, we outline the performance of the fully connected neural network used to train the Hammett-Perkins closure in the previous section for a data point which is generated from a slightly different underlying distribution and the result is shown in Figure \ref{fig:HP_4}. It can be seen that the true profile has more high frequency content which the trained network struggles to predict even though low frequency trends are recovered somewhat. The PDF of the true dataset is also recovered approximately (note how this new data set is distinctly multi-modal). The scatter plots also show consistent errors through the entirety of the domain.

\begin{figure}[!htb]
    \centering
    \includegraphics[width=0.8\textwidth]{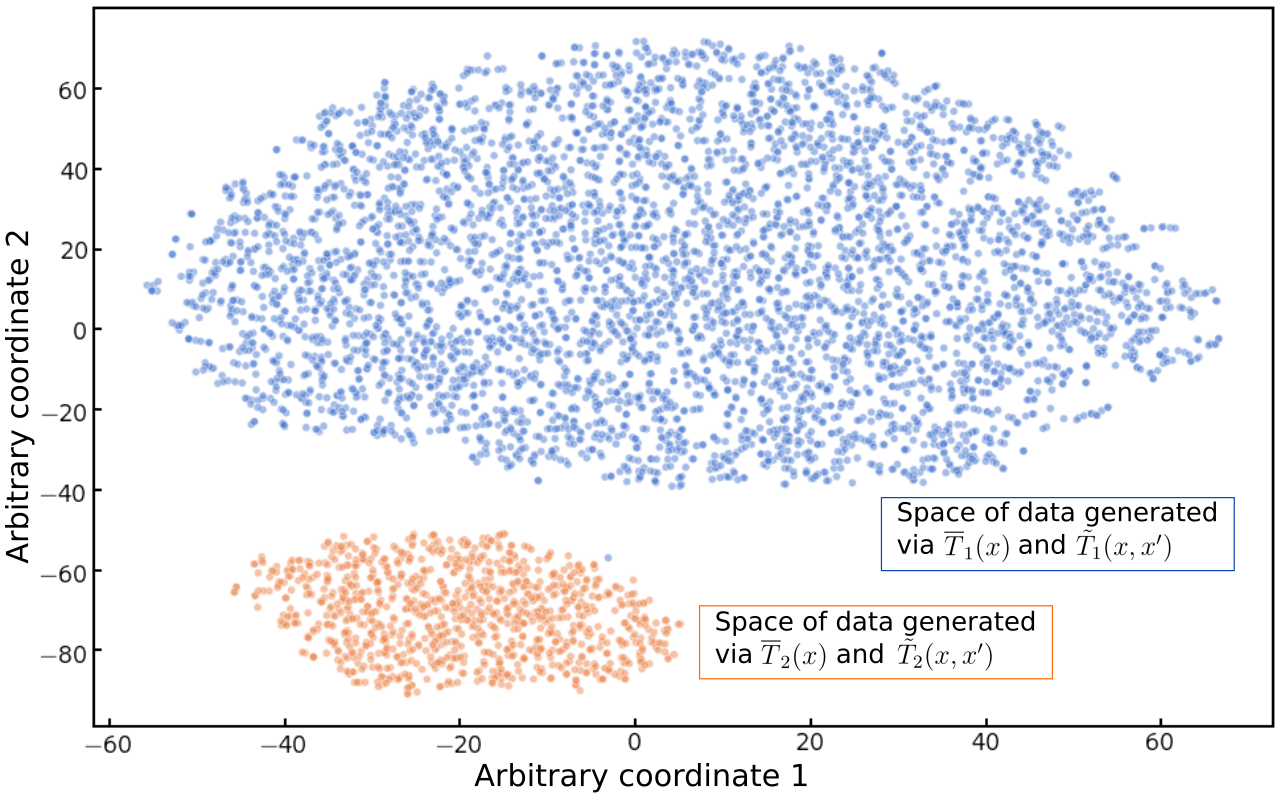}
    \caption{T-SNE encoding of training and testing datasets from different underlying distributions. The two clusters indicate two distinct distributions between training and testing. Formation of distinct clusters in input data is a key indicator
    of operating in an extrapolating fashion, in which care must be taken.}
    \label{fig:HP_5}
\end{figure}

Finally, we outline the use of T-distributed Stochastic Neighbor Embedding (t-SNE) \cite{hinton2003stochastic} which is a nonlinear embedding of our training and testing data onto a two-dimensional space. The embedding only requires the input data (i.e., the true values need not be known) and can be a check to determine the risk of extrapolation. Figure \ref{fig:HP_5} shows this analysis on the new dataset with the multimodal distribution of outputs which leads to the presence of two distinct clusters corresponding to the two distinct distributions in our total dataset. Note however, that this method cannot be coupled with a computational deployment of a closure as it is significantly costly in comparison with a forward pass through any of these networks. However, it can be a useful diagnostic to determine if a machine learning framework has contributed to failure of a simulation.

\section{Conclusion}\label{sec:five}
\begin{figure}[!htb]
    \centering
    \includegraphics[width=0.8\textwidth]{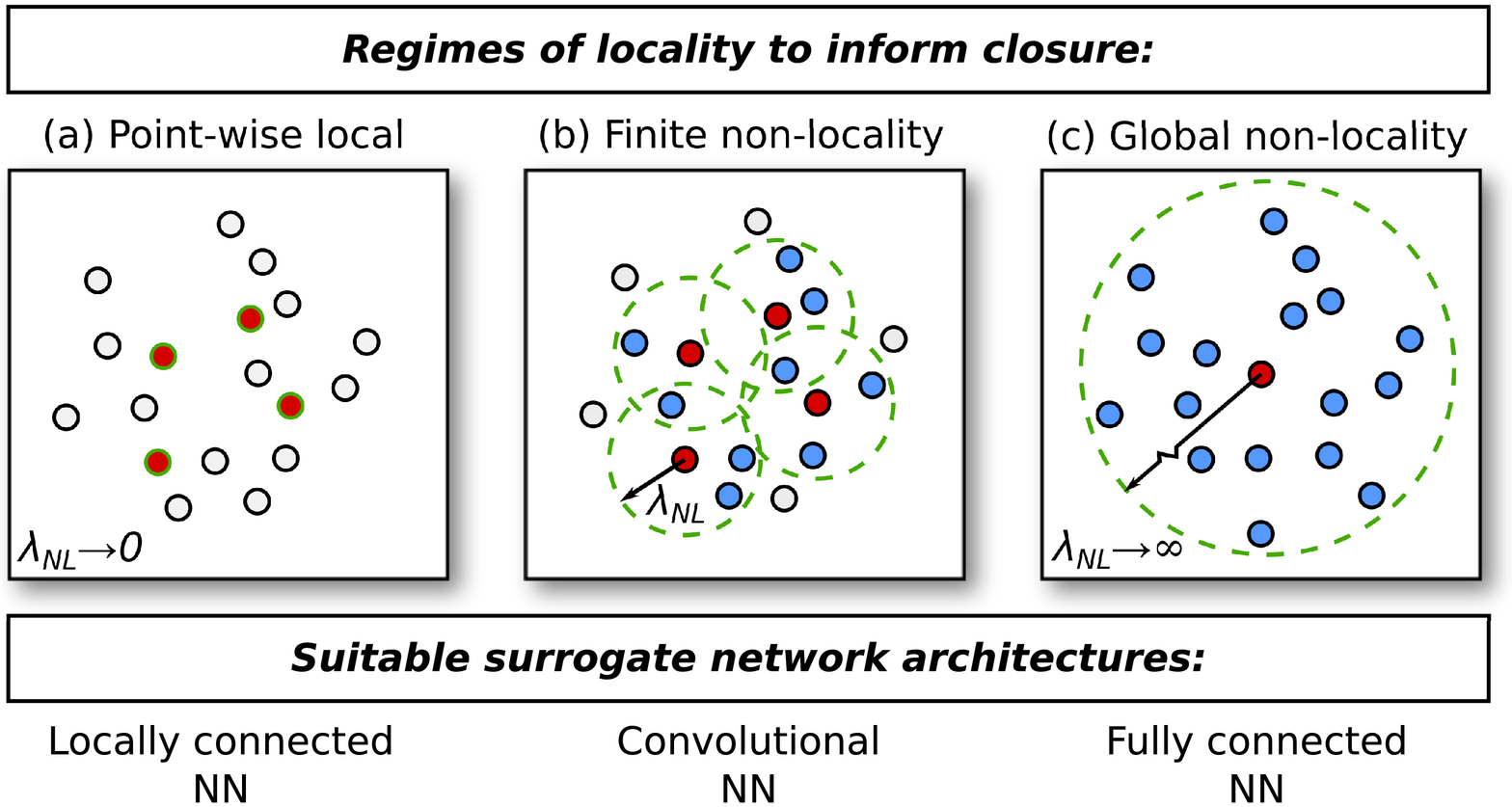}
    \caption{Simple schematic of varying classes of closure formulations. In each figure a red focus point can be thought to have a non-local interaction length, $\lambda_{NL}$, bounded by the green region containing blue points that influence the dynamics of a given focus point. This interaction region can be qualified in each region as (a) $\lambda_{NL}\rightarrow0$ and local information is sufficient, (b) finite $\lambda_{NL}$ and finite region of interaction, and (c) $\lambda_{NL}\rightarrow\infty$ and the entire global domain must be considered.}
    \label{fig:nn_summary}
\end{figure}

From this study, we find that in different physical regimes, an informed choice can be made on the neural network architecture chosen to formulate the closure surrogate model based on the physics expected in the model system. A simple schematic summary of this is shown Fig.~\ref{fig:nn_summary}, demonstrating that for a given knowledge of the locality of the system, one can choose a network architecture that is suitable to the physics inherent in the problem.

In the simplest limit is local closure, where point-wise values of closed quantities can be expressed as a functions of macroscopic values at that exact location in space. The primary example of this regime used in this work was the Braginskii closure for a collisional magnetized plasma, in which equilibration through collisions enables a local closure to be formulated for the variables that require closure. In this scenario, we observe that a simple single layer network can robustly and efficiently reproduce the required physics in high dimensional simulations without excessive computational burden that a more general connected network architecture would produce \cite{Ma_arxiv_2019}.

Next, one can also consider intermediate regimes where non-locality exists but is contained over a finite range of influence. This concept has previously been identified by Hazeltine \cite{hazeltine1998} in the context of source and boundary edge properties influencing the dynamics of the bulk plasma. In this scenario, we observe a CNN architecture, similar to that commonly used in image processing where there is a relationship between neighbouring points, yields a good balance between reproducing the required non-local physics, without a fully connected multilayered network.

Finally, in the extreme of a globally non-local closure, such as in the case in which collisionless-phase-mixing \cite{hammettphys.rev.lett.1990,HammettPoFB1992,TsiklauriPoP2008,Ma_arxiv_2019} provides a truly global influence on point-wise values, we find that a fully connected multilayered deep neural network is required to emulate the global non-locality of the closure.

\begin{acknowledgments}
We acknowledge productive discussions with Dr. Sandeep Madireddy and Dr. Bethany Lusch for this article. This material is based upon work supported by the U.S. Department of Energy (DOE), Office of Science, Office of Advanced Scientific Computing Research, under Contract~DE-AC02-06CH11357. This research was funded in part and used resources of the Argonne Leadership Computing Facility, which is a DOE Office of Science User Facility supported under Contract DE-AC02-06CH11357. Part of this work was performed at Los Alamos National Laboratory~(contract~No.~89233218CNA000001) and jointly supported by the Theory Program of the Office of Fusion Energy Sciences, and two SciDAC projects on Tokamak Disruption Simulation (TDS) and runaway electrons (SCREAM) by Office of Fusion Energy Science and Office of Advanced Scientific Computing.
\end{acknowledgments}

\nocite{*}
\bibliography{main.bib}

\end{document}